\documentclass[pre,showpacs,twocolumn]{revtex4}
\usepackage{graphicx}

\begin{document}

\title{Dynamics of neural cryptography}
\date{21 December 2006}
\author{Andreas Ruttor}
\author{Wolfgang Kinzel}
\affiliation{Institut f\"ur Theoretische Physik, Universit\"at
  W\"urzburg, Am Hubland, 97074 W\"urzburg, Germany}
\author{Ido Kanter}
\affiliation{Minerva Center and Department of Physics, Bar Ilan
  University, Ramat Gan 52900, Israel}
\begin{abstract}
  Synchronization of neural networks has been used for novel public
  channel protocols in cryptography. In the case of tree parity
  machines the dynamics of both bidirectional synchronization and
  unidirectional learning is driven by attractive and repulsive
  stochastic forces. Thus it can be described well by a random walk
  model for the overlap between participating neural networks. For
  that purpose transition probabilities and scaling laws for the step
  sizes are derived analytically. Both these calculations as well as
  numerical simulations show that bidirectional interaction leads to
  full synchronization on average. In contrast, successful learning is
  only possible by means of fluctuations. Consequently,
  synchronization is much faster than learning, which is essential for
  the security of the neural key-exchange protocol. However, this
  qualitative difference between bidirectional and unidirectional
  interaction vanishes if tree parity machines with more than three
  hidden units are used, so that those neural networks are not
  suitable for neural cryptography. In addition, the effective number
  of keys which can be generated by the neural key-exchange protocol
  is calculated using the entropy of the weight distribution. As this
  quantity increases exponentially with the system size, brute-force
  attacks on neural cryptography can easily be made unfeasible.
\end{abstract}
\maketitle

\section{Introduction}
\label{sec:intro}

Synchronization of neural networks \cite{Metzler:2000:INN,
  Kinzel:2003:DGI} is a special case of an online learning situation.
Two neural networks start with randomly chosen uncorrelated weights.
In each time step they receive common input values, communicate their
output to each other and use a suitable learning rule to update their
weights. Finally, this process leads to full synchronization of
corresponding weights in both networks.

In the case of simple networks, e.g., perceptrons, there is no
difference between unidirectional learning and bidirectional
synchronization. However, for tree parity machines (TPMs) an
interesting phenomenon can be observed: two neural networks learning
from each other synchronize much faster than a third network only
listening to the communication \cite{Kinzel:2003:DGI}.

This effect has been applied to solve a cryptographic problem
\cite{Kanter:2002:SEI}: Two partners $A$ and $B$ want to exchange a
secret message over a public channel. In order to protect the content
against an attacker $E$, who is listening to the communication, $A$
encrypts the message. However, then $B$ needs $A$'s key for
decryption.  Without an additional private channel $A$ and $B$ have to
use a cryptographic key-exchange protocol in order to generate a
common secret key over the public channel \cite{Stinson:1995:CTP}.
This can be achieved by synchronizing two TPMs, one for $A$ and one
for $B$, respectively. Of course, the attacker tries to determine the
key, too.  But when learning is much slower than synchronization, a
tree parity machine (TPM) trained by $E$ is usually unable to
synchronize before $A$ and $B$ have finished the key exchange.
Therefore the success probability $P_E$ of an attack is very small
\cite{Mislovaty:2002:SKE}.

Compared to other key-exchange algorithms neural cryptography needs
only simple mathematical operations, namely, adding and subtracting
integer numbers. Thus it is possible to use this key-exchange protocol
in devices with limited computing power. Computer scientists are
already working on hardware implementations, which are part of an
integrated circuit \cite{Volkmer:2004:LCS, Volkmer:2005:KEI,
  Volkmer:2005:LKE, Volkmer:2005:TPM}.

Since the first proposal of the neural key-exchange protocol
\cite{Kanter:2002:SEI} most research has been focused on finding more
advanced methods for the partners \cite{Mislovaty:2003:PCC,
  Ruttor:2004:NCF, Ruttor:2005:NCQ} and the attacker
\cite{Klimov:2003:ANC, Shacham:2003:CAN, Ruttor:2006:GAN}. However,
the results of simulations and iterative calculations show the same
scaling behavior in almost all cases: the success probability $P_E$
decreases exponentially with increasing synaptic depth $L$
\cite{Mislovaty:2002:SKE}, while the average synchronization time
$t_\mathrm{sync}$ only grows proportional to $L^2$
\cite{Ruttor:2004:SRW}. Therefore $L$ plays the same role in neural
cryptography as the key length in traditional cryptographic systems,
which are based on number theory \cite{Ruttor:2006:GAN}.

In this paper we analyze the synchronization process of two tree
parity machines by the dynamics of the overlap $\rho$. First, we
repeat the definition of basic algorithms of neural cryptography
regarding synchronization and attacks in Sec.~\ref{sec:neurocrypt}.
In Sec.~\ref{sec:frequency} we calculate the probabilities of
attractive and repulsive steps for different types of interactions.
The effect of these steps on the overlap is then presented in
Sec.~\ref{sec:dynamics}. Here we show that the mechanisms for
unidirectional learning and bidirectional synchronization are indeed
different. In Sec.~\ref{sec:entropy} we finally apply our results on
the dynamics of neural synchronization in order to analyze the
security of neural cryptography against brute-force attacks. For that
purpose we use the entropy of the weight distribution to determine a
scaling law for the number of keys which can be generated by the
neural key-exchange protocol.

\section{Neural synchronization}
\label{sec:neurocrypt}

\begin{figure}
  \centering
  \includegraphics[width=8.6cm]{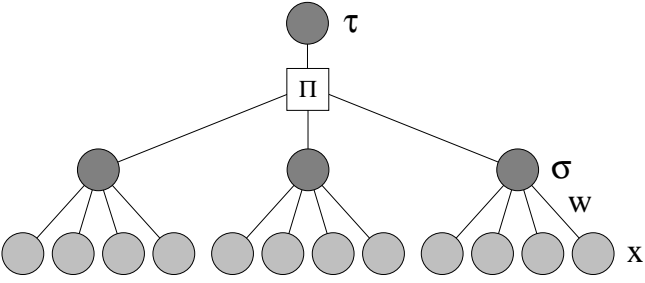}
  \caption{A TPM with $K=3$ and $N=4$.}
  \label{fig:tpm}
\end{figure}

The TPMs used by partners and attackers in neural cryptography consist
of $K$ hidden units, which are discrete perceptrons with independent
receptive fields. The general structure of these networks is shown in
Fig.~\ref{fig:tpm}. All input values are binary,
\begin{equation}
  x_{ij} \in \{-1,+1\}
\end{equation}
and the weights are discrete numbers between $-L$ and $+L$,
\begin{equation}
  w_{ij} \in \{-L,-L+1,\dots,+L\} \,.
\end{equation}
Here the index $i = 1,\dots,K$ denotes the $i$th hidden unit of the
TPM and $j = 1,\dots,N$ the elements of each vector.  As usual, the
output $\sigma_i$ of a hidden unit is given by the sign of the scalar
product of inputs and weights
\begin{equation}
  \sigma_i = \mathrm{sgn}(\mathbf{w}_i \cdot \mathbf{x}_i)
\end{equation}
and the total output $\tau$ of a TPM is defined as the product
(parity) of the hidden units
\begin{equation}
  \tau = \prod_{i=1}^{K} \sigma_i \,.
\end{equation}

The two partners start with secret random weight vectors
$\mathbf{w}^A$ and $\mathbf{w}^B$, respectively. At each time step, a
common public input vector $\mathbf{x}$ is generated, and the partners
exchange their output bits over the public channel. The weight vectors
are updated, and the process is iterated until the partners have
synchronized their weights which then are used for the secret key.
Note that the hidden units $\sigma_i^A$ and $\sigma_i^B$ are secret,
this is an essential mechanism for the security of neural
cryptography.

We consider three different algorithms for the update of the weights
in each time step.

\emph{Synchronization:}
\begin{equation}
  \label{eq:bdsync}
  \mathbf{w}_i^{A/B+} = \mathbf{w}_i^{A/B} + \mathbf{x}_i \,
  \Theta(\sigma_i^{A/B} \tau^A) \Theta(\tau^A \tau^B) \,.
\end{equation}
In neural cryptography this algorithm is used by the partners $A$ and
$B$. Here we only consider the random walk learning rule
\cite{Kinzel:2002:NC}, because all other suitable learning rules
(Hebbian and Anti-Hebbian) converge to it in the limit $N \rightarrow
\infty$ \cite{Ruttor:2006:GAN}.

\emph{Simple attack:}
\begin{equation}
  \label{eq:udsync}
  \mathbf{w}_i^{E+} = \mathbf{w}_i^E + \mathbf{x}_i \,
  \Theta(\sigma_i^E \tau^A) \Theta(\tau^A \tau^B) \,.
\end{equation}
This method is the simplest algorithm for unidirectional learning.  An
attacker $E$ can try it in order to synchronize with the partners $A$
and $B$ by training a TPM with the observed examples consisting of
$\mathbf{x}_i$ and $\tau^A$ \cite{Kanter:2002:SEI}.

\emph{Geometric attack:} The geometric attack is the most successful
method for an attacker using only a single TPM \cite{Klimov:2003:ANC}.
Here $E$ tries to realize Eq.~(\ref{eq:bdsync}) without being able to
interact with $A$. As long as $\tau^E = \tau^A$, this can be achieved
by just applying Eq.~(\ref{eq:udsync}), as both learning rules have
the same effect. However, in the case $\tau^E \neq \tau^A$ $E$ cannot
stop $A$'s update of the weights. Instead of this the attacker uses
additional information contained in the local fields
\begin{equation}
  h_i^E = \frac{1}{\sqrt{N}} \mathbf{w}_i^E \cdot \mathbf{x}_i
\end{equation}
of the hidden units in order to correct the output $\tau^E$ of her
TPM. As a low absolute value $|h_i^E|$ indicates a high probability of
$\sigma_i^E \neq \sigma_i^A$, the attacker flips the output of the
hidden unit with minimal $|h_i^E|$ before applying the learning rule
(\ref{eq:udsync}).

In all three cases weights $w_{ij}$ leaving the allowed range between
$-L$ and $+L$ are reset to the nearest boundary value
$\mathrm{sgn}(w_{ij}) L$.

We analyze the process of synchronization using simulations of finite
systems as well as iterative calculations for $N \rightarrow \infty$
\cite{Rosen-Zvi:2002:MLT, Ruttor:2004:NCF}. Correlations between the
weight vectors of two corresponding hidden units $i$ are described by
$(2 L + 1)^2$ variables $p_{a,b}^i(t)$, which are defined as the
probability to find a weight with $w_{ij}^A(t)=a$ in $A$'s tree parity
machine and $w_{ij}^B=b$ in $B$'s TPM at time $t$:
\begin{equation}
  p_{a,b}^i(t) = \mathrm{P}(w_{ij}^A(t)=a \wedge w_{ij}^B(t)=b) \,.
\end{equation}
While these quantities are approximately given by the frequency of the
weight values $w_{ij}^A(t)$ and $w_{ij}^B(t)$ in simulations, we use
the equations of motion given in Ref.~\cite{Ruttor:2004:NCF} to
determine the time evolution of $p_{a,b}^i(t)$ directly in the limit
$N \rightarrow \infty$.

In both cases the standard order parameters \cite{Engel:2001:SML},
which are commonly used for the analysis of online learning, can be
calculated as functions of $p_{a,b}^i(t)$:
\begin{eqnarray}
  Q_i^A =& \displaystyle \frac{1}{N} \mathbf{w}_i^A \cdot
  \mathbf{w}_i^A
  &= \sum_{a=-L}^{L} \sum_{b=-L}^{L} a^2 \, p_{a,b}^i \, ,\\
  Q_i^B =& \displaystyle \frac{1}{N} \mathbf{w}_i^B \cdot
  \mathbf{w}_i^B
  &= \sum_{a=-L}^{L} \sum_{b=-L}^{L} b^2 \, p_{a,b}^i \, ,\\
  R_i^{A,B} =& \displaystyle \frac{1}{N} \mathbf{w}_i^A \cdot
  \mathbf{w}_i^B
  &= \sum_{a=-L}^{L} \sum_{b=-L}^{L} a \, b \, p_{a,b}^i \,.
\end{eqnarray}
The level of synchronization between two corresponding hidden units is
then given by the normalized overlap \cite{Engel:2001:SML}
\begin{equation}
  \rho_i = \frac{\mathbf{w}_i^A \cdot
    \mathbf{w}_i^B}{\sqrt{\mathbf{w}_i^A \cdot \mathbf{w}_i^A}
    \sqrt{\mathbf{w}_i^B \cdot \mathbf{w}_i^B}} =
  \frac{R_i^{A,B}}{\sqrt{Q_i^A Q_i^B}} \,.
\end{equation}
Uncorrelated weight vectors at the beginning of the synchronization
process have $\rho_i=0$, while the maximum value $\rho_i=1$ is reached
for fully synchronized weights.

\begin{figure}
  \centering
  \includegraphics[width=8.6cm]{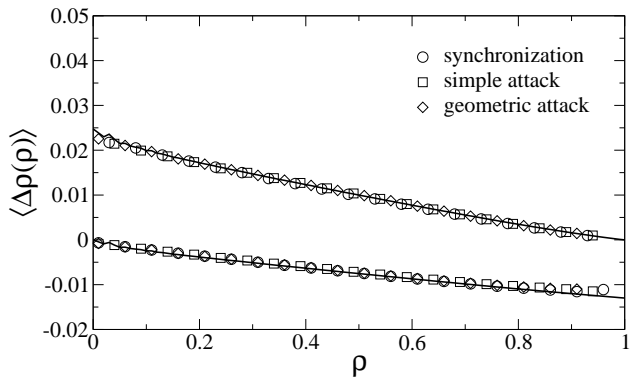}
  \caption{Effect of attractive (upper curve) and repulsive steps
    (lower curve) for $K=3$ and $L=10$. Symbols represent averages
    over $1000$ simulations for $N=100$. The lines denote the
    corresponding results obtained by iterative calculations for
    bidirectional synchronization and $N \rightarrow \infty$.}
  \label{fig:effect}
\end{figure}

All the update algorithms discussed above can be described by
\begin{equation}
  \mathbf{w}_i^{A/B/E}(t+1) = \mathbf{w}_i^{A/B/E}(t) + f_i^{A/B/E}(t)
  \mathbf{x}_i(t) \,,
\end{equation}
where $f^{A/B/E}(t)$ is a function, which can take the values $-1$,
$0$, or $+1$ according to the learning rule. Therefore only three
different effects are possible.

If $f_i^A(t) = f_i^B(t) \neq 0$, the weights in both corresponding
hidden units are moved in the same direction, so that the overlap
increases. This is called an ``attractive step'' and can be described
as anisotropic diffusion
\begin{equation}
  \label{eq:plus}
  p_{a,b}^{i+} = \frac{1}{2} \left( p^i_{a+1,b+1} + p^i_{a-1,b-1}
  \right) \, ,
\end{equation}
with reflecting boundary conditions. A sequence of these attractive
steps finally reaches a fixed point at $\rho_i=1$.

If only the weights of one hidden unit are updated, $f_i^A(t) +
f_i^B(t) = \pm 1$, the overlap $\rho_i$ decreases on average. This
repulsive step performs a normal diffusion on a $(2L + 1) \times (2L +
1)$ square lattice
\begin{equation}
  \label{eq:minus}
  p_{a,b}^{i+} = \frac{1}{4} \left( p^i_{a+1,b} + p^i_{a-1,b} +
    p^i_{a,b+1} + p^i_{a,b-1}\right) \,.
\end{equation}
Of course, the boundary conditions are the same as above. For a
sequence of these repulsive steps the fixed point of the overlap is
located at $\rho_i=0$.

For $f_i^A(t) = f_i^B(t) = 0$ the weights stay at their position.
Therefore the overlap does not change at all in this step.

The remaining situation $f_i^A(t) = -f_i^B(t) \neq 0$ cannot occur in
any algorithm discussed above.

In general, the change of the overlap $\Delta \rho$ is not only a
function of the current overlap, but depends also on the distribution
of the weights and the type of step, which is denoted by a subscript
if necessary: the effect of an attractive step is given by $\Delta
\rho_a(\rho)$, while we use $\Delta \rho_r(\rho)$ in the case of
repulsive steps. Both quantities as well as $\Delta \rho(\rho)$, which
is not restricted to a particular type of step, are random variables,
whose properties can be determined in simulations or iterative
calculations.

\begin{figure}
  \centering
  \includegraphics[width=8.6cm]{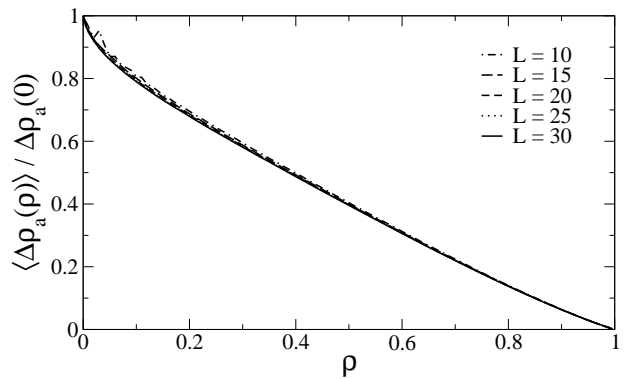}
  \caption{Scaling functions for attractive steps. These results were
    obtained in $1000$ iterative calculations for bidirectional
    synchronization with $K=3$ and $N \rightarrow \infty$.}
  \label{fig:atscaling}
\end{figure}

However, for special cases an analytical solution can be given by
using Eqs.~(\ref{eq:plus}) and (\ref{eq:minus}), taking the boundary
conditions into account. At the beginning of the synchronization all
weights are uniformly distributed, so that a repulsive step does not
change the overlap, but an attractive step has a large effect:
\begin{eqnarray}
  \label{eq:effect_a}
  \Delta \rho_a(\rho=0) &=& \frac{12 L}{(L + 1)(2 L + 1)^2} \sim
  \frac{3}{L^2} \, , \\
  \Delta \rho_r(\rho=0) &=& 0 \,.
\end{eqnarray}
In contrast, one observes the opposite situation for fully
synchronized weights ($\rho=1$):
\begin{eqnarray}
  \Delta \rho_a(\rho=1) &=& 0 \, , \\
  \label{eq:effect_r}
  \Delta \rho_r(\rho=1) &=& - \frac{3}{(L + 1)(2 L + 1)} \sim
  - \frac{3}{2 L^2} \,.
\end{eqnarray}
Here an attractive step does not change the overlap at all, but a
repulsive step has its maximum effect.

Figure~\ref{fig:effect} shows that $\langle \Delta \rho_a(\rho)
\rangle$ and $\langle \Delta \rho_r(\rho) \rangle$ do not depend on
the synchronization algorithm. Consequently, the difference between
unidirectional learning and bidirectional synchronization is caused by
the probability of attractive and repulsive steps, but not their
effects.

\begin{figure}
  \centering
  \includegraphics[width=8.6cm]{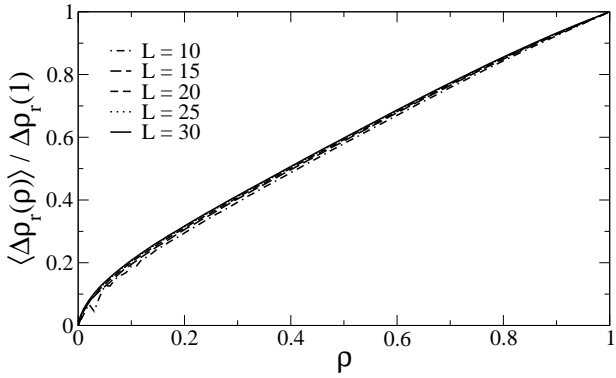}
  \caption{Scaling functions for repulsive steps. These results were
    obtained in the same way as those shown in
    Fig.~\ref{fig:atscaling}.}
  \label{fig:rpscaling}
\end{figure}

Using Eqs.~(\ref{eq:effect_a}) and (\ref{eq:effect_r}) we obtain the
rescaled quantities $\langle \Delta \rho_\mathrm{a}(\rho) \rangle /
\Delta \rho_\mathrm{a}(0)$ and $\langle \Delta \rho_\mathrm{r}(\rho)
\rangle / \Delta \rho_\mathrm{r}(1)$ which become asymptotically
independent of $L$ for large synaptic depth. This is clearly visible
in Figs.~\ref{fig:atscaling} and \ref{fig:rpscaling}. Therefore these
two functions are sufficient to describe the effect of attractive and
repulsive steps.

\section{Transition probabilities}
\label{sec:frequency}

All algorithms for neural synchronization have in common that a
repulsive step can only occur in the $i$th hidden unit, if the two
corresponding outputs $\sigma_i$ are different. The probability for
this event is given by the well-known generalization error
\cite{Engel:2001:SML}
\begin{equation}
  \label{eq:generr}
  \epsilon_i = \frac{1}{\pi} \mathrm{arccos}(\rho_i)
\end{equation}
of the perceptron. In the case of unidirectional learning mutual
interaction does not happen. Therefore the probability of repulsive
steps for an simple attack is directly given by Eq.~(\ref{eq:generr}):
\begin{equation}
  P_r^E = \epsilon_i \,.
\end{equation}
However, if both hidden units agree on $\sigma_i$, this does not
always lead to an attractive steps, because $\sigma_i = \tau$ is
another necessary condition for an update of the weights. Thus the
probability of attractive steps is given by
\begin{equation}
  P_a^E = \frac{1}{2} \left( 1 - \epsilon_i \right)
\end{equation}
in the case of learning with $K>1$.

In contrast, mutual interaction is an integral part of bidirectional
synchronization. When $\tau^A \neq \tau^B$, that move of the weights
would have a repulsive effect in at least one hidden unit, hence the
partners $A$ and $B$ avoid it by not updating the weights. However,
when an even number of hidden units disagrees on the output, one has
$\tau^A = \tau^B$ and the learning rule is applied. Taking all
possibilities into account, we find for $K=3$ and identical overlap
($\epsilon_i = \epsilon$) in all hidden units \cite{Ruttor:2004:NCF}:
\begin{eqnarray}
  P_a^B &=& \frac{1}{2} \frac{(1 - \epsilon)^3 + (1 -
    \epsilon) \epsilon^2}{(1 - \epsilon)^3 + 3 (1 - \epsilon)
    \epsilon^2} \, , \\
  P_r^B &=& \frac{2 (1 - \epsilon) \epsilon^2}{(1 -
    \epsilon)^3 + 3 (1 - \epsilon) \epsilon^2} \, .
\end{eqnarray}
Because of $P_r^B \leq P_r^E$ the partners have a clear advantage over
a simple attacker in neural cryptography.

\begin{figure}
  \centering
  \includegraphics[width=8.6cm]{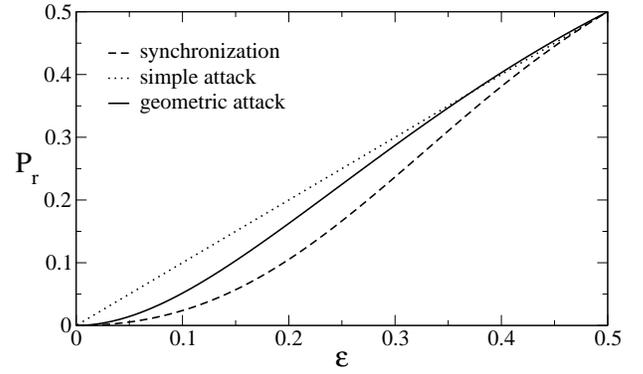}
  \caption{Probability that updating the weights has a repulsive
    effect in one hidden unit for $K=3$.}
  \label{fig:prcmp}
\end{figure}

However, $E$ can do better by taking the local field into account.
Then the probability for $\sigma_i^E \neq \sigma_i^A$ is given by the
prediction error \cite{Ein-Dor:1999:CPN}
\begin{equation}
  \label{eq:prediction}
  \epsilon^p_i = \frac{1}{2} \left[ 1 - \mathrm{erf} \left(
      \frac{\rho_i}{\sqrt{2(1-\rho_i^2)}} \frac{|h_i|}{\sqrt{Q_i}}
    \right) \right]
\end{equation}
of a perceptron, which depends not only on $\rho_i$ but also on
$|h_i^E|$. This quantity is a strictly monotonic decreasing function
of $|h_i^E|$. Therefore the geometric attack is often able to find the
hidden unit with $\sigma_i^E \neq \sigma_i^A$ by searching for the
minimum of $|h_i^E|$. If all other units have $\sigma_j^E =
\sigma_j^A$, then the probability for a successful geometric
correction \cite{Ruttor:2006:GAN} is given by
\begin{eqnarray}
  \label{eq:pg}
  P_g &=& \int_{0}^{\infty} \prod_{j \neq i} \left(
    \int_{h_i}^{\infty} \frac{2}{\sqrt{2 \pi Q_j}} \frac{1 -
      \epsilon^p_j}{1 - \epsilon_j} \, e^{-\frac{h_j^2}{2
        Q_j}} \, dh_j \right) \nonumber \\
  &\times& \frac{2}{\sqrt{2 \pi Q_i}}
  \frac{\epsilon^p_i}{\epsilon_i} \, e^{-\frac{h_i^2}{2 Q_i}}
  \, dh_i \,.
\end{eqnarray}
Using this equation, we find for $K=3$, geometric attack and identical
overlap in all hidden units
\begin{eqnarray}
  P_a^E &=& \frac{1}{2} (1 + 2 P_g) (1 - \epsilon)^2
  \epsilon + \frac{1}{2} (1 - \epsilon)^3 \nonumber \\
  &+& \frac{1}{2} (1 - \epsilon) \epsilon^2 + \frac{1}{6} \epsilon^3
  \, , \\
  P_r^E &=& 2 (1 - P_g) (1 - \epsilon)^2 \epsilon +
  2 (1 - \epsilon) \epsilon^2 + \frac{2}{3} \epsilon^3 \,.
\end{eqnarray}
While $P_r^E$ for the geometric attack is lower than for the simple
attack, it is still higher than $P_r^B$. Thus even this advanced
algorithm for unidirectional learning has a disadvantage compared to
bidirectional synchronization, which is clearly visible in
Fig.~\ref{fig:prcmp}.

\section{Dynamics of the overlap}
\label{sec:dynamics}

\begin{figure}
  \centering
  \includegraphics[width=8.6cm]{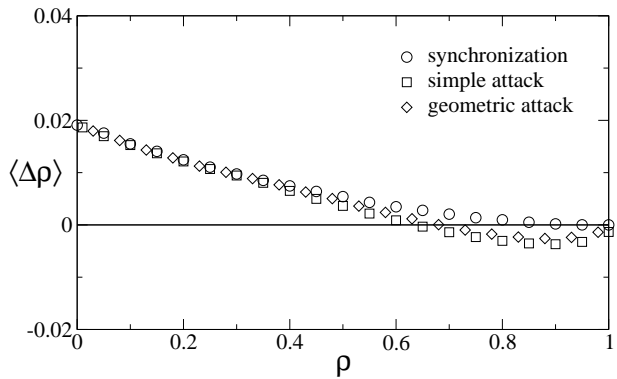}
  \caption{Average change of the overlap for $K=3$, $L=5$, and
    $N=100$. Symbols represent results obtained in $1000$
    simulations.}
  \label{fig:average}
\end{figure}

The results presented in Secs.~\ref{sec:neurocrypt} and
\ref{sec:frequency} indicate that the overlap $\rho$ between two
corresponding hidden units performs a random walk with position
dependent step sizes ($\Delta \rho_a(\rho)$, $\Delta \rho_r(\rho)$)
and transition probabilities ($P_a(\rho)$, $P_r(\rho)$). In order to
understand the dynamics, we calculate the average change of the
overlap as a function of $\rho$ itself:
\begin{equation}
  \langle \Delta \rho \rangle = P_a \Delta \rho_a +
  P_r \Delta \rho_r \,.
\end{equation}
This quantity is shown in Fig.~\ref{fig:average}. In the case of
bidirectional synchronization for $K \leq 3$ it is always positive
until the process reaches an absorbing state at $\rho = 1$.

While the transition probabilities are independent of $L$, the step
sizes decrease asymptotically proportional to $L^{-2}$ according to
Eqs.~(\ref{eq:effect_a}) and (\ref{eq:effect_r}). That is why $\langle
\Delta \rho \rangle$ is also proportional to $L^{-2}$ and we find
\cite{Mislovaty:2002:SKE}
\begin{equation}
  \label{eq:synctime}
  t_\mathrm{sync} \propto \frac{1}{\langle \Delta \rho \rangle}
  \propto L^2
\end{equation}
for the average number of steps needed for full synchronization. In
fact, the probability $P_\mathrm{sync}(t)$ to achieve identical weight
vectors in $A$'s and $B$'s neural networks in at most $t$ steps is
given by a Gumbel distribution
\begin{equation}
  \label{eq:gumbel}
  P_\mathrm{sync}(t) = e^{-e^\frac{\alpha-t}{\beta}} \, ,
\end{equation}
with parameters $\alpha$ and $\beta$, which increase proportional to
$L^2$ \cite{Ruttor:2004:SRW}. Thus $A$ and $B$ can generate a common
key in a short time with the help of neural cryptography.

Additionally, $\langle \Delta \rho(\rho) \rangle$ is independent of
$N$ except for finite-size effects, which is clearly visible in
Fig.~\ref{fig:effect}. The exact analytical calculation presented in
Ref.~\cite{Ruttor:2004:SRW} yields $t_\mathrm{sync} \propto L^2 \ln
N$, because one has to wait until all corresponding weights have
identical values. Consequently, the partners are able to use rather
large neural networks without problems.

However, the situation is completely different in the case of
learning, e.g., for the simple attack or the geometric attack. Now
there exists a fixed point of the dynamics at $\rho_f < 1$ with
$\langle \Delta \rho(\rho_f) \rangle = 0$. In the case of $K=3$, which
is the usual choice in regard to neural cryptography, we find $\rho_f
\approx 0.65$ for the simple attack and $\rho_f \approx 0.68$ for the
geometric attack.

\begin{figure}
  \centering
  \includegraphics[width=8.6cm]{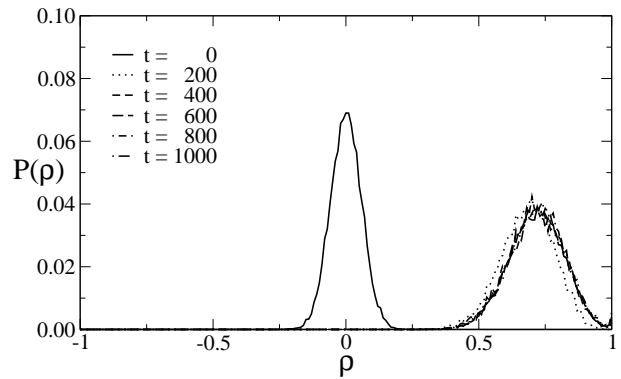}
  \caption{Distribution of the overlap $\rho$ at different time steps
    for $K=3$, $L=5$, $N=100$, and geometric attack. These histograms
    show the result of $10 \, 000$ simulations.}
  \label{fig:dist}
\end{figure}

As long as $\rho < \rho_f$ the overlap increases on average, but
afterwards we observe a quasistationary Gaussian distribution of
$\rho$ with mean value $\rho_f$ and standard deviation $\sigma_f$.
This is clearly visible in Fig.~\ref{fig:dist}. Consequently, the
absorbing state $\rho=1$ can only be reached by fluctuations of the
overlap.

In order to determine the scaling of the standard deviation $\sigma_f$
of the overlap at the fixed point, we use a linear approximation for
the dynamics of $\Delta \rho$ around $\rho_f$ without taking boundary
conditions into account
\begin{equation}
  \label{eq:diff}
  \Delta \rho(t) \approx - \gamma [\rho(t) - \rho_f] +
  \delta \xi(t) \,.
\end{equation}
Here $\xi(t)$ are random numbers with zero mean and unit variance. The
parameters are defined as
\begin{eqnarray}
  \gamma &=& - \left. \frac{d}{d \rho} \langle
    \Delta \rho(\rho) \rangle \right|_{\rho=\rho_f} \, , \\
  \delta &=& \sqrt{\langle [\Delta \rho(\rho_f)]^2 \rangle}
  \,.
\end{eqnarray}
In this model, the time evolution of the overlap is given as the
solution of Eq.~(\ref{eq:diff})
\begin{equation}
  \rho(t) - \rho_f = \sum_{i=1}^{t} (1 - \gamma)^{t - i}
  \delta \xi(t)
\end{equation}
using the initial condition $\rho(0) = \rho_f$, which is irrelevant in
the limit $t \rightarrow \infty$. Therefore the fluctuations of the
overlap in the stationary state are given by
\begin{equation}
  \sigma_\mathrm{f}^2 = \sum_{t=0}^{\infty} (1 - \gamma)^{2 t} \delta^2
  = \frac{\delta^2}{2 \gamma - \gamma^2} \,.
\end{equation}

\begin{figure}
  \centering
  \includegraphics[width=8.6cm]{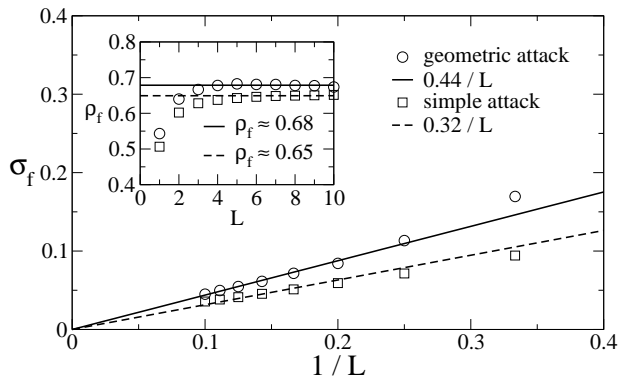}
  \caption{Standard deviation of $\rho$ at the fixed point
    $\rho_\mathrm{f}$. Symbols denote results averaged over $10 \,
    000$ simulations using $K=3$, $N=1000$, and unidirectional
    synchronization.}
  \label{fig:width}
\end{figure}

\begin{figure}
  \centering
  \includegraphics[width=8.6cm]{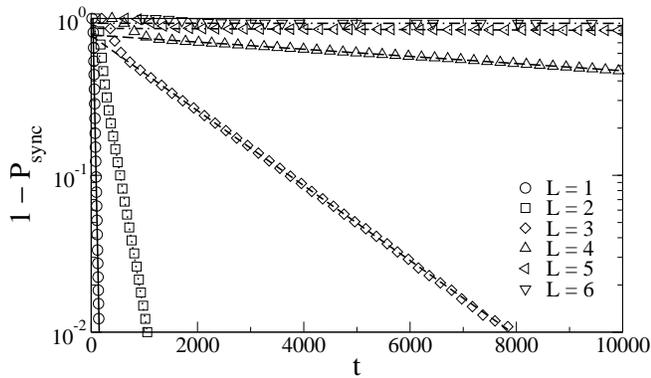}
  \caption{Probability distribution of $t_\mathrm{sync}$ for $K=3$,
    $N=1000$, and geometric attack. Symbols denote results averaged
    over $1000$ simulations and the lines show fits with
    Eq.~(\ref{eq:tdist}).}
  \label{fig:tdist}
\end{figure}

\begin{figure}
  \centering
  \includegraphics[width=8.6cm]{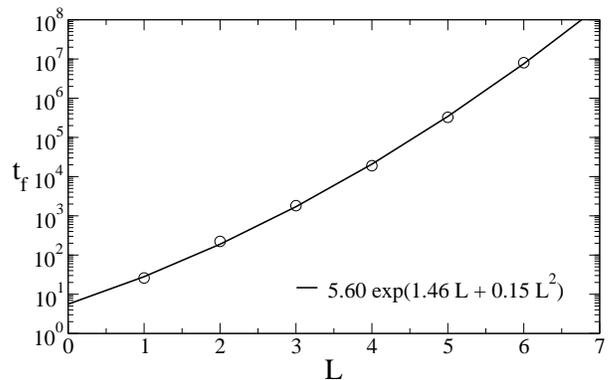}
  \caption{Time constant $t_\mathrm{f}$ for synchronization by
    fluctuations, obtained from $1000$ simulations of the geometric
    attack for $K=3$ and $N=1000$. The line shows a fit with
    Eq.~(\ref{eq:tf}).}
  \label{fig:time}
\end{figure}

\begin{figure}
  \centering
  \includegraphics[width=8.6cm]{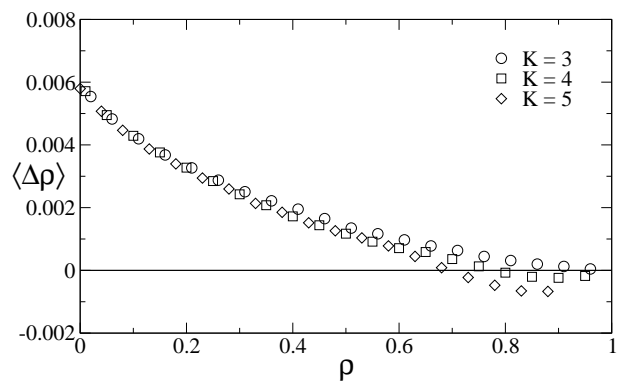}
  \caption{Average change of the overlap for bidirectional
    synchronization, $L=10$, and $N=1000$. Symbols denote results
    obtained from $100$ simulations.}
  \label{fig:kfour}
\end{figure}

As the step sizes of the random walk in $\rho$-space decrease
proportional to $L^{-2}$ for $L \gg 1$ according to
Eqs.~(\ref{eq:effect_a}) and (\ref{eq:effect_r}), this is also the
scaling behavior of the parameters $\gamma$ and $\delta$. Thus we find
\begin{equation}
  \sigma_f \propto \frac{1}{L}
\end{equation}
for larger values of the synaptic depth. Although we have neglected
the more complex features of $\langle \Delta \rho(\rho) \rangle$, this
scaling behavior is clearly visible in Fig.~\ref{fig:width}. The
deviations for small values of $L$ are caused by finite-size effects.

Consequently, an attacker $E$ is unable to synchronize with $A$ and
$B$ in the limit $L \rightarrow \infty$, even if she uses the
geometric attack. This is also true for any other algorithm, which has
a fixed point at $\rho_f < 1$ in the dynamics of the overlap.

For finite synaptic depth, however, $E$ has a chance of getting beyond
the fixed point at $\rho_f$ by fluctuations. The probability that this
event occurs in any given step is independent of $t$ once the
quasistationary state has been reached. That is why
$P^E_\mathrm{sync}(t)$ is no longer given by Eq.~(\ref{eq:gumbel}),
but described well for $t \gg t_0$ by an exponential distribution
\begin{equation}
  \label{eq:tdist}
  P^E_\mathrm{sync}(t) = 1 - e^{-\frac{t - t_0}{t_f}} \, ,
\end{equation}
with time constant $t_f$. This is clearly visible in
Fig.~\ref{fig:tdist}. Because of $t_f \gg t_0$ one needs
\begin{equation}
  \langle t_\mathrm{sync} \rangle \approx t_f
\end{equation}
steps on average to achieve full synchronization by unidirectional
learning.

In our simplified model with linear $\langle \Delta \rho(\rho)
\rangle$ the average time needed to reach $\rho=1$ starting at the
fixed point is given by
\begin{equation}
  t_f \approx \frac{1}{P(\rho=1)} = \sqrt{2 \pi} \,
  \sigma_f \, e^{\frac{(1 - \rho_f)^2}{2
      \sigma_f^2}}
\end{equation}
in the case of small fluctuations $\sigma_f \ll 1 - \rho_f$, where we
can assume that the distribution of $\rho$ is not influenced by the
presence of the absorbing state at $\rho=1$. Hence we expect
\begin{equation}
  t_f \propto e^{c L^2}
\end{equation}
for the scaling of the time constant, as $\sigma_f$ changes
proportional to $L^{-1}$, while $\rho_f$ stays nearly constant. And
Fig.~\ref{fig:time} shows that indeed $t_f$ grows exponentially with
increasing synaptic depth
\begin{equation}
  \label{eq:tf}
  t_f \propto e^{c_1 L + c_2 L^2} \,.
\end{equation}

Thus the complexity of attacks on the neural key-exchange protocol can
be controlled by choosing $L$. Or if $E$'s effort stays constant, her
success probability drops exponentially with increasing synaptic
depth. This has been observed in the case of the geometric attack
\cite{Mislovaty:2003:PCC} and even for advanced methods
\cite{Ruttor:2005:NCQ, Ruttor:2006:GAN}. Consequently, $A$ and $B$ can
reach any desired level of security by increasing $L$, as the
complexity of a single key exchange grows only proportional to $L^2$.

However, this is not true for $K>3$. As shown in Fig.~\ref{fig:kfour},
a fixed point at $\rho_f<1$ appears in the case of bidirectional
synchronization, too. Therefore Eq.~(\ref{eq:synctime}) is not valid
any more and $t_\mathrm{sync}$ increases exponentially with $L$. That
is why TPMs with four and more hidden units cannot be used in the
neural key-exchange protocol.

\begin{figure}
  \centering
  \includegraphics[width=8.6cm]{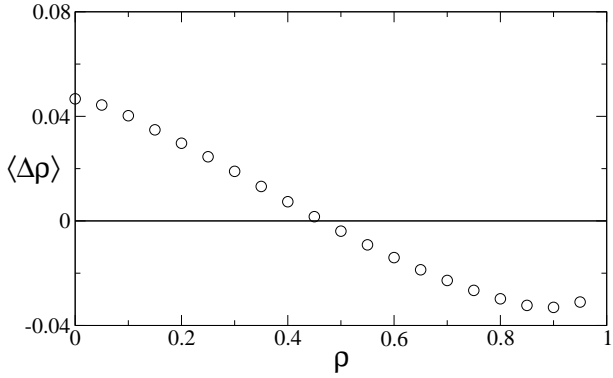}
  \caption{Average change of the overlap between $A$ and $C$ for
    $K=3$, $L=3$, and $N=1000$, obtained from $100$ simulations with
    $100$ pairs of TPMs.}
  \label{fig:kmove}
\end{figure}

\section{Number of keys}
\label{sec:entropy}

The state of the TPMs in each time step is a function of the secret
initial conditions and the public sequence of input values. Therefore
$E$ can---in principle---determine the possible weight configurations
at synchronization time $t_\mathrm{sync}$ using her knowledge about
the input vectors $\mathbf{x}_i(t)$. However, if the number of these
keys is large, a brute-force attack is unfeasible. Consequently, this
quantity is important for the security of neural cryptography, too.

In order to estimate the number of keys, which can be generated by the
neural key-exchange protocol using a given sequence of inputs, we look
at the following system consisting of two pairs of TPMs:
\begin{eqnarray}
  \mathbf{w}_i^{A+} &=& \mathbf{w}_i^A + \mathbf{x}_i
  \Theta(\sigma_i^A \tau^A) \Theta(\tau^A \tau^B) \, , \\
    \mathbf{w}_i^{B+} &=& \mathbf{w}_i^B + \mathbf{x}_i
  \Theta(\sigma_i^B \tau^B) \Theta(\tau^A \tau^B) \, , \\
  \mathbf{w}_i^{C+} &=& \mathbf{w}_i^C + \mathbf{x}_i
  \Theta(\sigma_i^C \tau^C) \Theta(\tau^C \tau^D) \, , \\
  \mathbf{w}_i^{D+} &=& \mathbf{w}_i^D + \mathbf{x}_i
  \Theta(\sigma_i^D \tau^D) \Theta(\tau^C \tau^D) \,.
\end{eqnarray}
In this model all four neural networks receive the same sequence of
inputs, but both pairs communicate their output bits only internally.
Thus $A$ and $B$ as well as $C$ and $D$ synchronize using the random
walk learning rule, while correlations caused by common inputs are
visible in the overlap $\rho_i^{AC}$. Because of the symmetry in this
system, $\rho_i^{AD}$, $\rho_i^{BC}$, and $\rho_i^{BD}$ have the same
properties as this quantity, so that it is sufficient to look at
$\rho_i^{AC}$ only.

Of course, synchronization of networks which do not interact with each
other, e.g., $A$ with $C$, is much more difficult and takes a longer
time than performing the normal key-exchange protocol. That is why we
assume $\rho^{AB}=1$ and $\rho^{CD}=1$ for the calculation of $\langle
\Delta \rho^{AC}(\rho^{AC}) \rangle$.

\begin{figure}
  \centering
  \includegraphics[width=8.6cm]{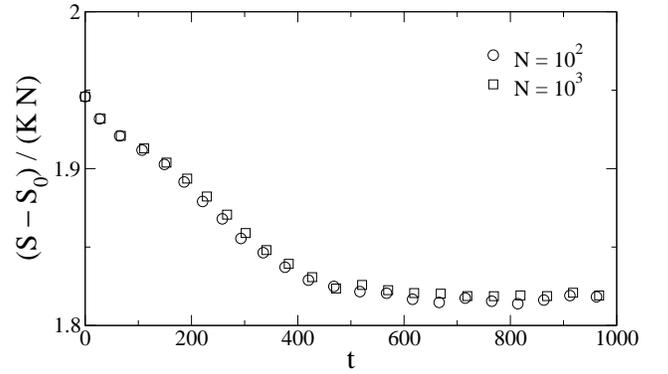}
  \caption{Entropy per weight for $A$ and $C$ for $K=3$ and $L=3$.
    Symbols denote results obtained in $100$ simulations with $100$
    pairs of TPMs.}
  \label{fig:entropy}
\end{figure}

The result is shown in Fig.~\ref{fig:kmove}. Similar to the case of
unidirectional learning, there is a fixed point at $\rho_f^{AC} < 1$
for the dynamics of the overlap. Because of $\rho_f^{AC} <
\rho_f^{AE}$ the probability for full synchronization in this case is
much smaller than for a successful simple attack. In fact, large
fluctuations which lead to equal weights without interaction only
occur in small systems. But the common input sequence causes
correlations between $\mathbf{w}_i^A$ and $\mathbf{w}_i^C$ even for $L
\gg 1$ and $N \gg 1$. Consequently, the number of keys
$n_\mathrm{key}$ is smaller than the number of weight configurations
$n_\mathrm{conf} = (2 L + 1)^{K N}$ of a tree parity machine.

We further analyze these correlations by calculating the entropy of
the weight distribution
\begin{equation}
  S = - K N \sum_{a=-L}^{L} \sum_{c=-L}^{L} p_{a,c} \ln p_{a,c} \,.
\end{equation}
Here $p_{a,c}$ is the probability to find $w_{ij}^A=a$ and
$w_{ij}^C=c$ by selecting a random weight. As the weights in each tree
parity machine alone stay uniformly distributed, the entropy of two
fully synchronized networks is given by
\begin{equation}
  \label{eq:s0}
  S_0 = \ln n_\mathrm{conf} = K N \ln (2 L + 1) \, ,
\end{equation}
which is also the entropy of a single TPM.  Consequently, the quantity
$S - S_0$ describes the correlations between the weight vectors caused
by the common input sequence. It is proportional to the length of the
generated cryptographic key with any redundancy removed using a
suitable encoding. Therefore the logarithm of the effective number of
keys is given by
\begin{equation}
  \label{eq:keys}
  \ln n_\mathrm{key} = S - S_0 \,.
\end{equation}
We note, however, that the real number can be larger, because not all
possible weight configurations occur with equal probability as keys.
Therefore $n_\mathrm{key}$ is, in fact, a lower bound for the number
of different final configurations. However, this quantity determines
the security against brute-force attacks, as an attacker tries the
most probable keys first.

\begin{figure}
  \centering
  \includegraphics[width=8.6cm]{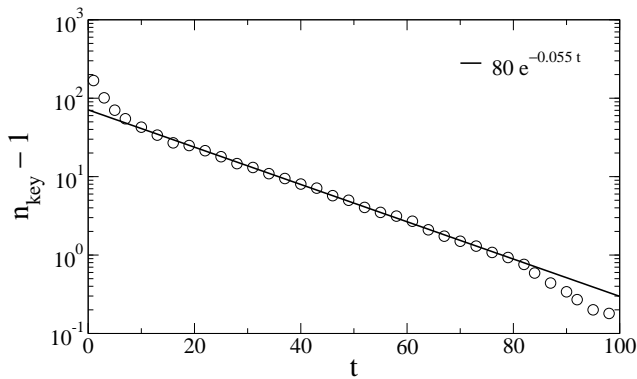}
  \caption{Number of keys for $K=3$, $L=1$, and $N=2$, obtained by
    exhaustive search and averaged over $100$ random input sequences.}
  \label{fig:keys}
\end{figure}

Figure~\ref{fig:entropy} shows the time evolution of this entropy.
First $S - S_0$ shrinks linearly with increasing $t$, as the overlap
$\rho$ between $A$ and $C$ grows while it approaches the stationary
state. This behavior is consistent with an exponential decreasing
number of keys, which can be directly observed in very small systems
as shown in Fig.~\ref{fig:keys}. Of course, after the system has
reached the fixed point shown in Fig.~\ref{fig:kmove}, the entropy
stays constant. We use this minimum value in order to determine
$n_\mathrm{key}$.

It is clearly visible that there are two scaling relations for $S(t)$.
The synchronization time $t_\mathrm{sync} \propto L^2$ is the time
scale of all processes related to the synchronization of tree parity
machines. It depends on the size of the learning steps $\langle \Delta
\rho \rangle$. Therefore the time needed to reach the fixed point of
$\rho_k^{AC}$ is proportional to $L^2$, too.

Entropy is an extensive quantity. Thus $S$ and $S_0$ are proportional
to the number of weights $N$. Consequently the number of keys, which
can be generated by the neural key-exchange protocol for a given input
sequence, grows exponentially with increasing $N$.

In order to determine the dependency between the synaptic depth $L$
and $n_\mathrm{key}$ we calculate the mutual information
\begin{equation}
  I = 2 S_0 - S
\end{equation}
between $A$ and $C$, which is visible in the correlations of the
weight vectors. Using Eqs.~(\ref{eq:s0}) and (\ref{eq:keys}) we find
\begin{equation}
  I = - \ln \left( \frac{n_\mathrm{key}}{n_\mathrm{conf}} \right) \,.
\end{equation}
Therefore the number of keys is given by
\begin{equation}
  n_\mathrm{key} = n_\mathrm{conf} \, e^{- I} = \left[ (2 L + 1)^K
    e^{- I / N} \right]^N \,.
\end{equation}

\begin{figure}
  \centering
  \includegraphics[width=8.6cm]{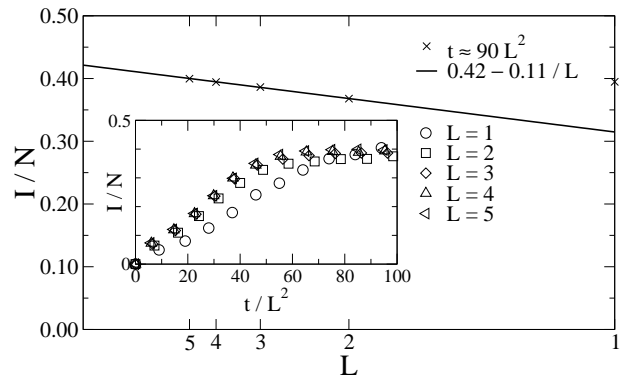}
  \caption{Mutual information between $A$ and $C$ for $K=3$, $N=1000$,
    obtained in $1000$ simulations with $10$ pairs of TPMs.}
  \label{fig:minfo}
\end{figure}

As shown in Fig.~\ref{fig:minfo}, $I / N$ becomes asymptotically
independent of the synaptic depth in the limit $L \rightarrow \infty$.
Of course, changing $N$ does not influence $I / N$ either, as it is an
intensive quantity. Extrapolating $I$ yields the result
\begin{equation}
  I \approx 0.42 N \,,
\end{equation}
which is valid for $K=3$ and $L \gg 1$. Consequently, $n_\mathrm{key}$
increases exponentially with $N$,
\begin{equation}
  n_\mathrm{key} \approx \left[ 0.66 (2 L + 1)^3 \right]^N \,,
\end{equation}
so that there are always enough possible keys in larger systems to
prevent successful brute-force attacks on the neural key-exchange
protocol.

\section{Conclusions}
\label{sec:cc}

Synchronization of neural networks is a dynamical process driven by
attractive and repulsive stochastic forces. While there is little
difference between unidirectional and bidirectional interaction in the
case of simple networks such as perceptrons, more complex networks
such as TPMs show an interesting phenomenon: neural networks which
interact with each other synchronize faster than those, which are only
trained with the examples generated by others.

We have investigated the dynamics of this effect, which is essential
for the recently proposed neural key-exchange protocol. In multilayer
feed-forward networks the hidden units are not public. Therefore
learning steps can have either an attractive or repulsive effect. In
both cases the step size only depends on the synaptic depth $L$ and
the time-dependent overlap $\rho$ between the networks. Two neural
networks, $A$ and $B$, learning from each other are able to skip
unsuitable input vectors because of their interaction. That is why
they avoid some repulsive steps and have a clear advantage over a
third passive neural network $E$, which cannot influence $A$ and $B$.
Consequently, $A$ and $B$ have a lower frequency of repulsive learning
steps than $E$, which causes the difference between bidirectional
synchronization and unidirectional learning.

Using the step sizes $\Delta \rho_a(\rho)$, $\Delta \rho_r(\rho)$ and
transition probabilities $P_a(\rho)$, $P_r(\rho)$ we described the
process of neural synchronization as a random walk of the overlap
$\rho$. The most important properties of the dynamics are visible in
the average change of the overlap $\langle \Delta \rho(\rho) \rangle$.
In the case of $K=3$ and bidirectional interaction the dynamics of the
overlap has only one fixed point at $\rho=1$. That is why full
synchronization is achieved after $\langle t_\mathrm{sync} \rangle
\propto L^2$ steps on average. However, for unidirectional learning or
mutual learning with $K>3$ there is an additional fixed point at
$\rho_f < 1$, so that $\rho=1$ is only reachable by fluctuations. This
leads to a different scaling behavior of the average synchronization
time $\langle t_\mathrm{sync} \rangle \propto e^{c_1 L + c_2 L^2}$.
Thus the difference between bidirectional synchronization and
unidirectional learning can be controlled by choosing the synaptic
depth $L$.

An identical input sequence causes correlations between tree parity
machines even without any other interaction. Similarly to the case of
unidirectional learning there is a fixed point at $\rho_f < 1$. As the
distance $1 - \rho_f$ is larger, full synchronization without
interaction is only observed for very small TPMs. But the correlations
restrict the number of different keys $n_\mathrm{key}$, which can be
generated by the neural key-exchange protocol using a certain input
sequence and random initial weights. Both the configuration space
$n_\mathrm{conf} = (2 L + 1)^{K N}$ and $n_\mathrm{key}$ grow
exponentially with increasing number of weights per hidden unit.
Therefore a large value of $N$ guarantees the security of neural
cryptography against brute-force attacks and similar methods.

\bibliography{paper}

\end{document}